# Coupling the thermal acoustic modes of a bubble to an optomechanical sensor

K. G. Scheuer[1], F. B. Romero[2], and R. G. DeCorby[2*]

We report experimental observations of the volume acoustic modes of air bubbles in water, including both the fundamental Minnaert breathing mode and a family of higher-order modes extending into the megahertz frequency range. Bubbles were placed on or near optomechanical sensors having a noise floor substantially determined by ambient medium fluctuations, and which are thus able to detect thermal motions of proximate objects. Bubble motions could be coupled to the sensor through both air (*i.e.*, with the sensor inside the bubble) and water, verifying that sound is radiated by the high-order modes. We also present evidence for elastic-Purcell-effect modifications of the sensor's vibrational spectrum when encapsulated by a bubble, in the form of cavity-modified linewidths and line shifts.

## Introduction

Any 'coherent' source of energy, such as an electric dipole or mechanical harmonic oscillator, is in a two-way communication with its environment. For example, according to the Fluctuation-Dissipation (FD) theorem [1], any damping or 'dissipation' mechanism in a system at thermal equilibrium must be accompanied (*i.e.*, in a statistical sense) by an equal and opposite forcing or 'fluctuation' mechanism. Mechanical oscillators [2,3], of particular interest here, are typically coupled to several thermal baths [4], and each of these baths must then also be viewed as a source of random thermal agitation reducing the coherence (*i.e.*, the quality factor, $Q$) of the oscillator.

If the environment is not homogeneous, its structure is imprinted on the fluctuating noise forces driving the oscillator. For example, the random pressure fluctuations in a gas are modified next to a hard boundary [1]. Related to this, it has been shown [5] that detailed information about an inhomogeneous environment can be extracted from correlations of the noise signals generated by one or more acoustic sensors. While this principle has primarily been applied to studies of the earth's crust using networks of low-frequency seismic detectors [6], it has also been verified at high ultrasonic frequencies using piezo-electric [5] or capacitive [7] sensors.

The FD theorem requires a thermodynamic (statistical) balance between the energy flowing to and from an oscillator; however, the coupling between an oscillator and its environment goes beyond that. Specifically, as first shown by Purcell [8], a structured environment modifies the behavior of a source by altering the spatial/spectral properties of the available radiation modes. As widely studied in the electromagnetics domain [9,10], both enhancement and suppression of spontaneous emission rates and changes to the peak emission frequency [11] are possible by placing a source (*e.g.*, an oscillating electric dipole) in a suitably engineered environment. Only recently, acoustic/elastic analogues of these effects have been demonstrated [12,13].

Here, we study the interactions between an optomechanical oscillator and an adjacent air bubble in water. Strong signatures of their interaction, driven only by room-temperature thermal energy, are passively recorded in the noise floor of the sensor. This reveals a family of bubble acoustic modes extending into the MHz-range, anticipated theoretically but not previously observed. Moreover, we provide strong evidence for Purcell-effect modification of the sensor's own vibrational spectrum due to altered acoustic density of states (DOS) within the bubble environment. To our knowledge, this is the first experimental demonstration of elastic Purcell effects at ultrasonic frequencies.

Correspondence: Ray DeCorby (rdecorby@ualberta.ca)
[1]Ultracoustics Technologies Ltd., Sherwood Park, AB, Canada, T8A 3H5
[2]ECE Department, University of Alberta, 9211-116 St. NW, Edmonton, AB, Canada, T6G 1H9

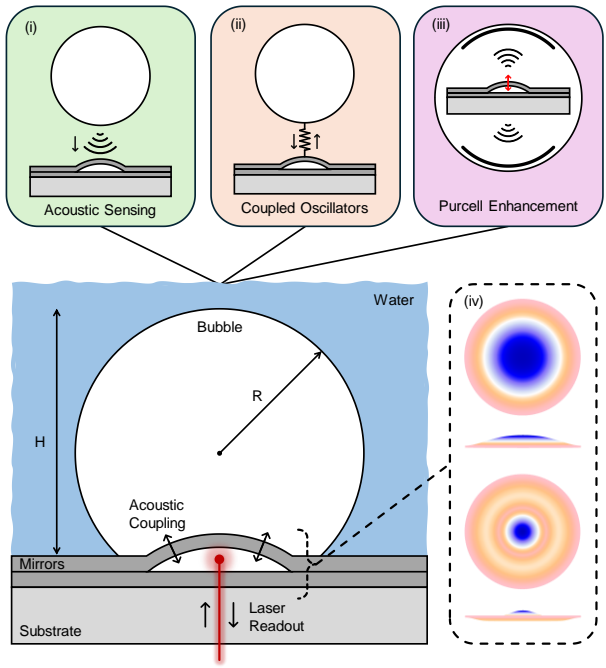

**Fig. 1. Experimental schematic and alternate viewpoints.** A schematic sketch of a tethered bubble aligned overtop of a Fabry-Perot sensor. The mechanical element in our sensor is a curved, buckled mirror, and its displacement is read out using a laser. In a simplistic view (i), acoustic energy confined in the bubble is detected as an external force on the sensor. A more holistic view (ii) treats the bubble and sensor as coupled oscillators. An alternative point of view (iii) treats the sensor as an oscillating 'dipole' source and the bubble as a cavity which modifies the acoustic density of states in the vicinity of the sensor. Panel (iv) shows overhead and side views of the two lowest-order, radially symmetric modes of the sensor's buckled mirror as predicted by a finite-element simulation (COMSOL).

## Experimental setup and sensor overview

Our sensors are buckled-dome, Fabry-Perot optomechanical cavities, described in detail elsewhere [14,15] and shown schematically in Fig. 1. The devices used here have a base diameter of 100 μm, a cavity length ~ 2.4 μm, and support a family of high-quality ($Q \sim 10^4$) Laguerre-Gaussian optical modes in the 1550 nm range. The buckled upper mirror functions as the mechanical oscillator, with its two lowest-order, radially symmetric vibrational modes in air centered at ~ 2.5 MHz and ~ 6 MHz, respectively (see Fig. 4 below). Numerically predicted (COMSOL) mode-field profiles for these are shown in panel (iv) of Fig. 1 and are analogous to those of a Chinese gong [12], only at ~ 5 orders of magnitude higher frequency.

Operation in a thermomechanical-noise-limited regime [16] is achieved for laser interrogation powers as low as ~ 10 – 100 μW. Moreover, ambient medium fluctuations make a significant contribution to the noise floor of these and other optomechanical ultrasound sensors [17], such that they are ideally suited to passive, noise-based [5] sensing of their environments. For the measurements described below, the reflected laser light was delivered to a high-speed photodetector and power spectral density (PSD) plots were generated from sampled noise signals. The laser power is sufficiently low (< 50 μW) such that it simply acts as a passive probe of the vibrational motion of the mirror while back-action effects are negligible [15]. Further details are provided in the supplementary information file.

## Bubble Acoustics

Bubbles host rich physics with important technological implications [18], especially regarding their interactions with acoustic waves [19]. Natural oscillations of entrained gas bubbles produce audible signals, such as the familiar sound of running water [19,20]. When they are actively driven by an external pressure wave, the cyclic collapse of bubbles can result in extremely energetic processes such as the cavitation-induced damage of solid objects [19]. Other phenomena associated with bubble cavitation include the emission of light, (sonoluminescence [21]) and the catalysis of reactions (sonochemistry [22]). Moreover, there is a growing interest in bubble-mediated nonlinear interactions between phonons and photons [23].

Many acoustic properties of bubbles can be explained in terms of the well-known Minnaert 'breathing mode' [18,20,24]. For a spherical air-bubble in water at atmospheric pressure, the resonant frequency ($f_M$) of this mode can be approximated from $f_M \cdot R \sim 3.3$ m s$^{-1}$ [25], where $R$ is the radius of the bubble. It follows that the associated acoustic wavelength (in both air and water) is much larger than the bubble dimensions (i.e., $\lambda_M \gg R$) [20]. In fact, the Minnaert breathing mode can be obtained by linearizing the so-called Rayleigh-Plesset (RP) equation [20,25], which (*a priori*) assumes that the pressure inside the bubble (i.e., in the air) is a spatially uniform function of time only.

Minnaert [24] used an energy balance argument to explain the origin of the audible-range sound produced by millimeter-scale bubbles. This phenomenon was not predicted by the earlier 'rigid acoustic sphere' model attributed to Paget (see Ref. [20] for a historical account) or by Lamb's capillary theory [26], which describes surface-tension-mediated 'shape modes' [19]. In spite of the fact that gas compression (rather than surface tension) provides its restoring force, the Minnaert breathing mode has often been characterized as the zero-order solution within the set of capillary modes [27-29].

Notwithstanding the shortcomings of Paget's 'rigid sphere' model, there is no doubt that a bubble can be viewed as an elastic body bounded by a viscous, compressible fluid, or approximately as a spherical resonant cavity [30-32]. Accordingly, a bubble also supports a set of *volume* acoustic modes, for which the pressure (and related parameters) inside the bubble are functions of both time *and position*. Taking this point of view, Devaud *et al.* [20] solved for a family of radial acoustic modes, drawing an analogy to the Fabry-Perot modes of a spherical-mirror optical cavity. By properly accounting for the compressibility and inertia of both the liquid and gas media, they demonstrated that the Minnaert resonance is in fact the lowest-order mode within this set. Its low resonant frequency (*i.e.*, a wavelength much larger than the bubble dimensions) was attributed to the dispersion imparted by the curved bubble interface. Thus, Devaud's analysis shows that the Minnaert breathing mode is more naturally aligned with Paget's original theory than it is with Lamb's theory.

While the higher-order acoustic modes have been considered in theoretical treatments of collapsing bubbles and cavitation [33-35], to date there is a scarcity of experimental evidence for their existence as 'natural resonances'. It is worth noting that the volume acoustic modes of liquid droplets have similarly only recently been observed [36-38].

To provide context for the results below, we consider first the archetypal case of a spherical air bubble in a water medium. As mentioned, Devaud *et al.* [20] provided a quasi-analytical derivation of the radially symmetric volume modes for this system. Inside the air bubble, these modes can be expressed as $p(r) = (A/r) \cdot \sin(q_a \cdot r)$, where $p$ is pressure, $r$ is the radial coordinate, $A$ is a constant amplitude, $q_a = (\omega/c)$ is the wave number, and $c$ is the sound velocity in air. They furthermore showed that the resonant frequencies for a bubble of radius $R$ are given by:

$$x_a = q_a R \approx 0.0623, 4.49, 7.73, \ldots \approx \frac{(2n+1)\pi}{2} \quad . \quad [1]$$

The lowest-order resonance is the well-known Minnaert breathing mode, characterized by a nearly homogeneous internal pressure. The others are higher-order, radially symmetric, volume acoustic modes. Notably, the first higher-order (radially symmetric) resonance is predicted to lie at ~ 72× higher frequency than the fundamental Minnaert resonance. As an example, for a typical millimeter-scale air bubble in water with Minnaert frequency $f_M$ ~ 10 kHz, this places the next radial resonance at ~ 720 kHz.

A simpler model would treat the bubble as a spherical acoustic resonator with hard boundaries [31,32], an approximation which can be justified by the large acoustic impedance mismatch between the air and the water. In this simple system, the air cavity hosts eigen-modes with pressure distribution given by:

$$p_{nlm}(r,\theta,\varphi) \sim \left\{ j_l\left(\frac{\omega_{nl} \cdot r}{c}\right) \right\} \cdot \{L_l^m(\cos\theta)\} \cdot \left\{ \begin{matrix} sin \\ cos \end{matrix}(m\varphi) \right\} , \quad [2]$$

where $\varphi$ and $\theta$ are the azimuthal and polar angles, respectively, $L_l^m$ is a Legendre function of the first kind (degree *l*, order *m*) [30], and $j_l$ is a spherical Bessel function. The resonance frequencies are given by $\omega_{nl} = z_{nl} \cdot c/R$, where $z_{nl}$ is the *n*th zero of the derivative of $j_l$, and the modes are degenerate for integer values of $m \leq |l|$. Solutions include a subset of purely radial modes:

$$p_n(r) \sim j_0\left(\frac{\omega_{n0} \cdot r}{c}\right) = \sin\left(\frac{\omega_{n0} \cdot r}{c}\right) / \left(\frac{\omega_{n0} \cdot r}{c}\right) , \quad [3]$$

where $\omega_{n0} = \{0, 4.49, 7.73, 10.90, 14.07\ldots\} \cdot c/R$ are the radial-mode eigen-frequencies in good agreement with Eq. (1). However, the two lowest-order (non-static) modes are non-radial, including the lowest-order mode:

$$p_{111}(r,\theta,\varphi) \sim \cos\varphi \cdot \sin\theta \cdot j_1\left(\frac{\omega_{11} \cdot r}{c}\right) , \quad [4]$$

with $\omega_{11} = 2.08 \cdot c/R$.

To further assess the validity of the hard-boundary approximation, we modeled the air-bubble-in-water system using the 'Pressure Acoustics' module in COMSOL Multiphysics. For simplicity, the air and water were assigned fixed sound velocities (340 and 1500 m/s, respectively) and absorption was neglected. Further details of the simulation, including boundary conditions, are provided in the supplementary information file.

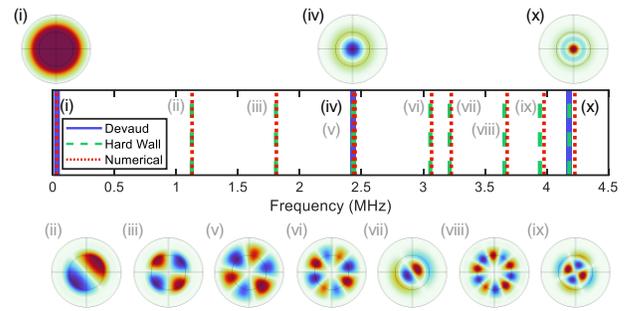

**Fig. 2. Acoustic modes for a spherical (non-tethered) bubble.** The predicted eigen-frequencies for the acoustic modes of a 100-μm-radius bubble are indicated by the vertical lines. Results from a numerical solver (COMSOL), an analytical bubble model for radial modes only [20], and an analytical model for a spherical resonator with hard walls [31] are indicated by the legend. The numerically predicted pressure distributions for the ten lowest-order modes are also shown, with the gas/liquid boundary indicated by the inner concentric circle. The mode labeled as '(i)' is the Minnaert breathing mode, and the others are higher-order acoustic modes.

The predictions of the analytical and numerical models are compared in Fig. 2. For the higher-order modes, there is nearly perfect agreement between all three models, including in the ordering and pressure profiles of the modes. Slight discrepancies in the eigen-frequencies predicted by the COMSOL model can be attributed in part to numerical error arising from a finite mesh size and simulation volume. Notably, the Minnaert resonance, which cannot be captured by the hard boundary model, is well-predicted by the other two models, supporting the contention [20] that it naturally fits within this larger set of volume acoustic modes.

## Results

### Bubble acoustic modes

We now turn our attention towards the experimental verification of the bubble acoustic modes. This was achieved by placing a tethered bubble over an optomechanical sensor, effectively positioning a sensor inside a bubble, as illustrated in Fig. 1. Sensor chips (~ 1 cm × 1 cm) were first entirely covered with a "puddle" of DI water and then air bubbles were injected using a syringe and needle [29]. Note that variations in water volume can impact the observed resonance frequencies [20], but was not studied here. A technique was developed for accurately positioning a bubble over an individual sensor, by first tethering it to the substrate plane and then dragging it with the dispensing needle. A video demonstrating this process is included as supplementary information. All experiments were performed at room temperature and in ambient laboratory conditions.

Results for a relatively small bubble ($R$ ~ 152 μm and $H$ ~ 266 μm) are shown in Fig. 3. The interrogation laser was coupled to a particular sensor, and sensor noise spectra were recorded at fixed laser power. The black curve in Fig. 3 is the background spectrum measured in "bulk" air (*i.e.*, not covered in liquid), and it is dominated by a typical [15] series of resonant peaks (*e.g.*, at ~ 2.5 and 6 MHz) associated with the inherent vibrational modes of the buckled mirror. The red curve is the spectrum measured with a small, tethered air bubble positioned over the sensor as shown in the inset.

Clearly, the noise spectrum is modified relative to the bulk case, most apparently by the appearance of several new resonant peaks. These peaks are well correlated with the four lowest-order acoustic-mode eigen-frequencies of the tethered bubble as predicted by a COMSOL numerical model and indicated by the dashed vertical lines. The correspondingly predicted spatial pressure distributions are also shown, evincing strong analogies with the higher-order modes solved for the spherical bubble case above. For the COMSOL model, the bubble dimensions were estimated from top- and side-view microscope images. Also, the chip surface (including the flexible buckled mirror) was set as a hard acoustic boundary. This is somewhat simplistic, and treats the bubble's acoustic/vibrational modes as completely independent from those of the sensor (*i.e.*, as depicted in panel (i) of Fig. 1). A more rigorous approach would consider the bubble and sensor as a pair of coupled harmonic oscillators [37] as depicted in panel (ii) of Fig. 1.

It is worth noting that the off-resonance displacement sensitivity of our devices is ~ $10^{-17}$ - $10^{-16}$ m/Hz$^{1/2}$ for operation in air [15]. Furthermore, a straightforward application of the equipartition theorem (see the SI file for further details) predicts that the air displacement associated with the bubble acoustic modes at room temperature exceeds this limit by at least an order of magnitude. In other words, the fact that these thermally driven modes appear as strong features is consistent with expectations.

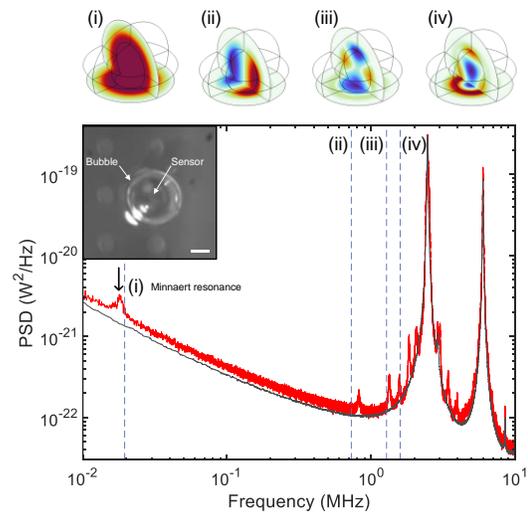

**Fig. 3. Experimental results for a small, tethered bubble. a.** Power spectral density noise plots for the same sensor in air (black curve, gently smoothed) and encapsulated by a bubble (red curve, with no smoothing). The vertical dashed lines indicate numerically predicted eigen-frequencies for the four lowest-order volume acoustic modes of the bubble. Corresponding acoustic pressure distributions (i-iv) are also shown. The innermost concentric circles are the air-water boundaries of the bubble, projected onto two orthogonal planes including the substrate boundary. The inset shows a top-down-view microscope image of the bubble tethered over the sensor of interest, and with several adjacent sensors visible (scale bar – 100 μm).

Slight discrepancies between the experimental and theoretical eigen-frequencies can be attributed to uncertainties in estimating bubble dimensions from the microscope images, and to the neglect of hybridization between sensor and bubble vibrational modes in the numerical model. Nevertheless, the global alignment is very good, and this was consistently observed across multiple bubble-sensor combinations (see Figs. 4 and 5 below and the SI file for additional examples), allowing us to confidently assert that these measurements are in fact revealing the high-frequency volume modes of the bubbles. Notably, the Minnaert resonance was also imprinted on the noise spectrum of the sensor in all cases, and the resonant frequency was consistently in excellent agreement with the COMSOL model and with analytical predictions [39] for a tethered bubble (see the SI file for additional discussion).

Analogous results for a larger tethered bubble are shown in Fig. 4, revealing, as expected, a denser spectrum of higher-order acoustic modes. From microscope images (Figs. 4(a) and (b)), we estimated $R \sim 313$ μm and $H \sim 526$ μm. The fundamental breathing (*i.e.*, Minnaert) resonance lies at $\sim 10$ kHz in this case, in good agreement with analytical [39] and numerical predictions, as shown in Fig. 4(c). In addition to the raw noise spectrum shown in Fig. 4(d), the normalized spectrum (*i.e.*, the PSD with bubble overtop divided by that for bulk air) is plotted in Fig. 4(e) to more clearly delineate the spectral features attributable to the bubble. As above, the vertical dashed lines indicate the six lowest-order eigen-frequencies as predicted by the finite element model and using the bubble dimensions estimated from microscope images. The predicted mode-field amplitude plots for these bubble modes are shown in Fig. 4(f).

Evidence of acoustic modes extending up to nearly 10 MHz is apparent, but assignment of individual modes becomes difficult due to the high density of modes predicted above $\sim 1$ MHz. The quality factor of the lower-order modes is as high as $\sim 70$ (see the SI file for further discussion), in good agreement with theoretical predictions considering radiation, thermal, and viscous damping [20]. We speculate that the lower quality of the modes above 1 MHz might in part be attributable to the dramatic rise of ultrasound attenuation in air [40].

Our data provides support to the viewpoint that the Minnaert resonance belongs to a larger set of acoustic modes mediated by compression [20]. Conversely, we saw no evidence of the low-frequency capillary modes [26-29] in our experiments. This is consistent with the fact that those surface-tension-mediated 'shape' modes are not efficiently coupled to acoustic radiation iufields [19,29].

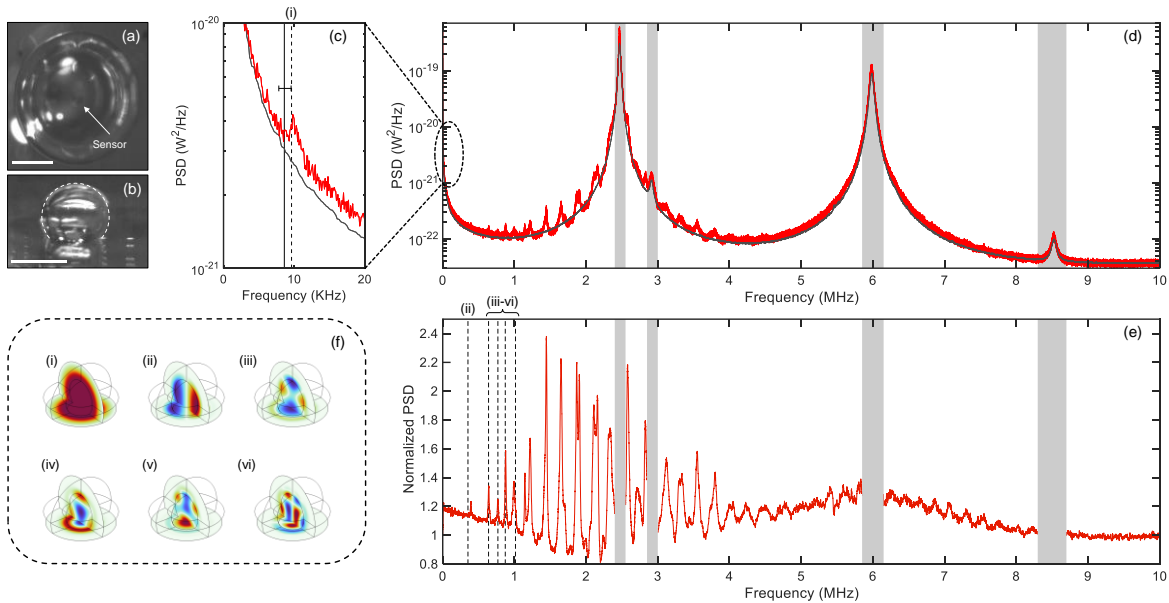

**Fig. 4. Analogous results to those presented in Fig. 3 but for a larger bubble. a,b.** Microscope images showing the top and side views of the bubble (scale bars – 200 μm and 500 μm, respectively). **c** Low-frequency-range PSD plots measured with the bubble overtop (red, not smoothed) and for bulk air (black, gently smoothed) revealing the Minnaert resonance at $\sim 10$ kHz. Analytical and numerical predictions of the Minnaert frequency are indicated by the solid and dashed vertical lines, respectively. The error bar represents a ± 10% deviation in the estimated bubble diameter. **d**. The raw noise spectrum measured with the bubble overtop (red, not smoothed) and for bulk air (black, gently smoothed). **e**. The PSD with bubble overtop the sensor normalized to that in bulk air. The next 5 numerically predicted eigen-frequencies are also plotted as vertical dashed lines. The grey bands represent regions dominated by mechanical modes inherent to the sensor. **f**. The predicted acoustic pressure distributions (COMSOL) for the lowest-order modes having the eigen-frequencies shown in parts c. and e.

We also studied cases where two tethered bubbles were located in the vicinity of a sensor of interest, as shown for example in Fig. 5. Note that the larger bubble in this case is centered overtop the sensor, as for the bubbles above, while the smaller bubble is tethered to the chip surface at an adjacent location. We observed clear signatures of acoustic coupling between the bubbles, which are manifested in the noise floor of the sensor. For example, a pair of resonances in the kHz frequency range are visible in this case, which we attribute to the hybridized Minnaert breathing modes of the two bubbles. Note that the simulated Minnaert frequency for the larger tethered bubble, shown by the dashed vertical line, lies between the two observed resonances, as would be expected due to mode hybridization.

Some evidence for coupling and hybridization of the higher-order acoustic modes can also be seen in the region above 1 MHz of Fig. 5. However, across multiple trials, this was less consistently observed than was the 'splitting' of the Minnaert resonance. Nevertheless, it does suggest that energy associated with the higher-order bubble modes is radiated into the surrounding water medium [20]. Further evidence for this was obtained from experiments in which the sensor of interest was not encapsulated by a bubble, while air bubbles were either tethered to the chip surface nearby or suspended from the needle in proximity to the sensor (see the SI file for these results). In both cases, signatures of MHz-range bubble acoustic modes were observed in the sensor noise spectrum. However, they were typically much lower in amplitude than for the bubble-encapsulated-sensor, since only a fraction of the energy circulating inside the bubble is radiated into the water [20].

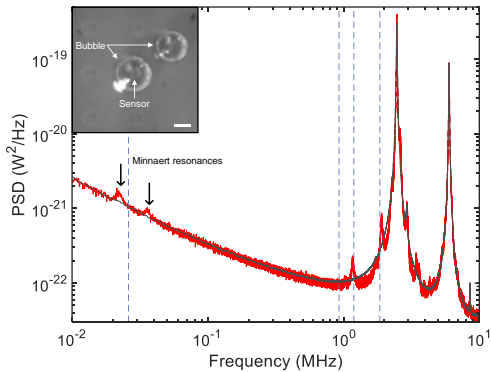

**Fig. 5. Experimental results for a sensor coupled to two bubbles.** Power spectral density plots for the same sensor in air (black, gently smoothed) and encapsulated by the larger bubble with the smaller tethered bubble nearby (red, not smoothed). The vertical dashed lines indicate numerically predicted eigen-frequencies for the lowest-order volume acoustic modes of the larger bubble but without accounting for the neighboring small bubble. The inset shows a top-down-view microscope image of the bubble pair and surrounding sensors (scale bar – 100 µm).

*Elastic Purcell effects*

So far, we have treated the optomechanical sensor as a passive detector of the resonant acoustic energy confined in an adjacent bubble. This approach is simplistic, but nevertheless provided clear evidence for the presence of the anticipated volume acoustic modes of the bubble. In this section, we explore the alternative point of view depicted in panel (iii) of Fig. 1, where the sensor is treated as a dipole (or similar) source and the bubble as an acoustic cavity. From this perspective the changes in the vibrational spectrum of the sensor can be attributed to changes in the acoustic DOS in its local environment.

Our system can be viewed as an elastic/acoustic analogue of the well-known experiment of Heinzen and Feld, in which they used a laser to probe atoms inside a confocal resonator [10,11]. Specifically, our 'emitter' is also coupled to three-dimensional cavity modes, and is also probed by a readout laser. Moreover, the coupling of our emitter to other loss channels, for example intrinsic flexural and clamping losses [4], is analogous to the atomic radiation out the 'sides' of their resonator. The representative results shown in Fig. 6 strongly support this analogy, as follows:

i. Especially for small bubbles, as shown in Fig. 6(a) and (d), we observed suppression of the sensor's vibrational motion over the ~ 0 – 8 MHz range, except at higher-$Q$ bubble resonances. Consistent with this, we observed large reductions (up to nearly 2×, see Fig. 6(a)) in the linewidth of the fundamental sensor resonance. This corresponds to an increase in its coherence (*i.e.*, emitter lifetime), which can be attributed to a reduction in the acoustic radiation loss inside the small bubble environment.

ii. We observed bubble-dependent line-shifts in the sensor's fundamental vibrational resonance, as shown in Figs. 6(a)-(c). Moreover, the direction of this shift depended on whether the nearest cavity (*i.e.*, bubble) mode was red- or blue-detuned relative to the sensor resonance. This is attributable to mode-coupling effects, which 'push' the sensor resonance away from the nearest cavity resonance [11].

iii. In cases where a cavity mode was aligned to the sensor mode, we observed significant enhancement (as high as ~ 2×) of the sensor's resonant vibrational energy, as shown for example in Fig. 6(b).

iv. Higher-order vibrational modes of the sensor were relatively unmodified, especially in larger bubbles (see the upper-left inset in Fig 6(c)), which can be attributed to the lack of isolated high-$Q$ bubble modes in their vicinity.

We observed similar behavior for multiple bubble/sensor combinations (see the SI file). These observations are consistent with a redistribution of the vibrational energy of the sensor, caused by the modified acoustic environment. The degree of modification is quite remarkable given that the mechanical oscillator is coupled to other thermal baths, in particular the underlying substrate. It also provides further evidence that the noise floor of these sensors is substantially limited by viscous damping and external radiation to their external medium [14,17].

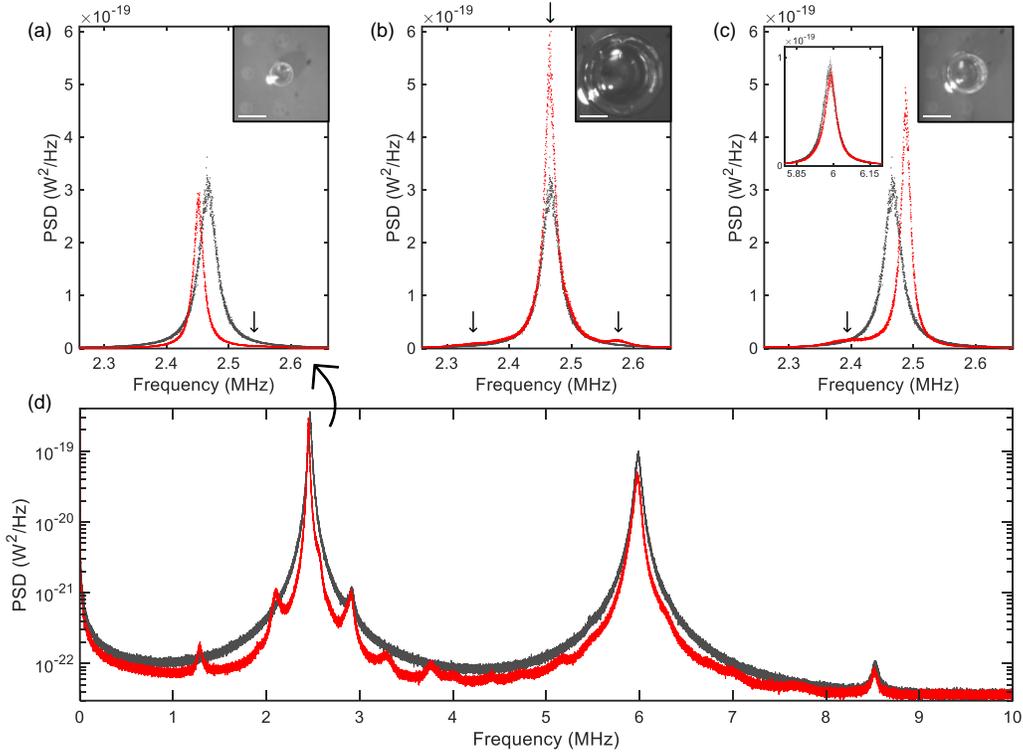

**Fig. 6. Evidence of elastic Purcell-effect modification of sensor vibrational modes. a-c.** Power spectral density noise plots in the vicinity of its fundamental resonance frequency for the same sensor in bulk air (black curves) and encapsulated by three different bubbles (red curves) shown in the upper-right inset microscope images (scale bars – 200 μm). The black vertical arrows indicate approximate locations of the nearby cavity (bubble) eigen-frequencies in each case. The upper-left inset in part c. shows the noise spectrum near the second resonance line of the sensor, which is relatively unmodified in this case due to the lack of isolated, nearby cavity modes. **d** A wider range noise plot for the small bubble from part a., showing evidence of suppressed vibrational energy extending up to ~ 8 MHz, except for enhancement in the vicinity of a few low-order bubble acoustic modes.

It should be noted that modified emission by an acoustic source in the vicinity of reflecting boundaries is well-known from a classical perspective [41]. However, interpretation of such phenomena in terms of Purcell-effects has only recently appeared in the literature [12,13,42,43]. A rigorous treatment requires calculation of the spatial emission pattern of the source [12], and its spatial and spectral overlap with the environmental modes of interest [13,42]. This is a complicated endeavor for our system, and is left for future work. Nevertheless, it is interesting to consider the 'ideal' elastic Purcell factor for an emitter perfectly aligned to a single cavity mode, $F_P \sim (3/4\pi^2) \cdot Q/(V/\lambda^3)$ [13,42], where $Q$ is the effective quality factor (taking into account both cavity and emitter linewidth), $V$ is the effective cavity mode volume, and $\lambda$ is the acoustic wavelength. Using $Q \sim 50$ and $V \sim 2 \cdot \lambda^3$ (*i.e.*, $\lambda \sim 140$ μm in air at ~ 2.5 MHz) yields $F_P \sim 1.9$, in reasonable agreement with the enhancement observed in Fig. 6(b).

## Discussion and Conclusions

In summary, we described an experimental study of coupling interactions between optomechanical sensors and air bubbles in water, for frequencies extending into the MHz ultrasound range. The results provide unique insights into the acoustic properties of bubbles, as well as compelling evidence for the Purcell-effect modification of a mechanical oscillator.

Regarding bubble acoustics, our results confirm that the vibrational properties of a bubble go beyond the Minnaert breathing mode and capillary 'shape' modes [19,24-29] and include a set of higher-order volume modes [20] not previously observed. It is fair to ask whether these latter modes are mainly of theoretical interest, or rather might have practical implications. Undoubtedly, the situations where they are expected to manifest are fewer than those for the Minnaert resonance, simply because of the increased attenuation of sound with frequency, particularly in air. Moreover, the Minnaert resonance is unique in the sense that it involves prominent motion of the relatively massive water medium, which ties it more directly to the dramatic effects associated with acoustic cavitation.

Nevertheless, the internal state and dynamics of a gas bubble is an incredibly complex physical problem, especially in scenarios involving the collapse of oscillating bubbles [25]. It seems plausible that a complete description might need to include consideration of the higher-order acoustic modes. Notably, energy storage by acoustic modes of a bubble has been posited as a potential contributor to the extreme conditions leading to single bubble sonoluminescence [33], although the same authors subsequently discounted this theory [21] due to a lack of experimental corroboration. Notwithstanding this point of view, the role of acoustic modes remains a matter of ongoing debate [34,45]. Moreover, bubble acoustics is a central theme in several emerging fields, including phonon-photon interactions mediated by bubbles [23] and the use of bubbles in biosensing and related applications [46,47].

Regarding Purcell effects, our results demonstrate that mesoscopic optomechanical oscillators are a uniquely accessible platform for such studies. Notably, a mechanical 'emitter' (*i.e.*, mechanical oscillator) can be more easily and directly probed [12] than the atomic emitters used in typical electromagnetics studies. For example, the behavior of an atomic emitter is often inferred indirectly from the spectral/spatial characteristics of the emitted photons, which makes it challenging to separate changes in the intrinsic behavior of the emitter, such as modified linewidth and Lamb shifts [11], from the classically predicted spatial redistribution of the emitted light [44].

In our experiments, on the other hand, the laser directly interrogates the motion of the buckled mirror, so that changes in the thermomechanical noise spectrum can be mapped directly to changes in its vibrational behavior. Accordingly, the results provide clear evidence for Purcell-effect modifications of a mesoscopic mechanical oscillator at MHz-range frequencies, including cavity-modified lineshifts and frequency-dependent suppression or enhancement of vibrational motion. We hope that these results might prompt further research at the intersection between quantum electrodynamics and optomechanics.

**Acknowledgements**
This research was funded by the Government of Alberta (Innovation Catalyst Grant), Alberta Innovates, the Natural Sciences and Engineering Research Council of Canada (CREATE 495446-17), and the Alberta EDT Major Innovation Fund (Quantum Technologies).

**Author contributions**
RGD led the conception and interpretation of the experiment, and the manuscript preparation. KGS led the design of the experiment, carried out the numerical simulations, and assisted with both the interpretation of results and manuscript preparation. FBR conducted the experiment and collected data, assisted by KGS.

**Conflicts of interest**
Ultracoustics Technologies Ltd. (I,P) KGS, North Road Photonics Corp. (I,P) RGD.

**Supplementary information**
The online version contains supplementary material available at:

**References**
1. H. B. Callen and T. A. Welton, "Irreversibility and generalized noise," Phys. Rev. **83**(1), 34-40 (1951).
2. T. B. Gabrielson, "Mechanical-thermal noise in micromachined acoustic and vibration sensors," IEEE Trans. Electron Devices **40**(5), 903-909 (1993).
3. B. D. Hauer, C. Doolin, K. S. D. Beach, and J. P. Davis, "A general procedure for thermomechanical calibration of nano/micro-mechanical resonators," Ann. Phys. **339**, 181-207 (2013).
4. M. Aspelmeyer, T. J. Kippenberg, and F. Marquardt, "Cavity optomechanics," Rev. Mod. Phys. **86**(4), 1391-1452 (2014).
5. R. L. Weaver and O. I. Lobkis, "Ultrasonics without a source: thermal fluctuation correlations at MHz frequencies," Phys. Rev. Lett. **87**, 134301 (2001).
6. N. M. Shapiro, M. Campillo, L. Stehly, and M. H. Ritzwoller, "High-resolution surface-wave tomography from ambient seismic noise," Science **307**, 1615-1618 (2005).
7. S. Lani, S. Satir, G. Gurun, K. G. Sabra, and F. L. Degertekin, "High frequency ultrasonic imaging using thermal mechanical noise recorded on capacitive micromachined transducer arrays," Appl. Phys. Lett. **99**, 224103 (2011).
8. E. M. Purcell, "Spontaneous emission probabilities at radio frequencies," Phys. Rev. **69**, 681 (1946).
9. S. Haroch and D. Kleppner, "Cavity quantum electrodynamics," Phys. Today **42**(1), 24-30 (1989).
10. D. J. Heinzen, J. J. Childs, J. E. Thomas, and M. S. Feld, "Enhanced and inhibited visible spontaneous emission by atoms in a confocal resonator," Phys. Rev. Lett. **58**(13), 1320–1323 (1987).
11. D. J. Heinzen and M. S. Feld, "Vacuum radiative level shift and spontaneous-emission linewidth of an atom in an optical resonator," Phys. Rev. Lett. **59**(23), 2623-2626 (1987).
12. L. Langguth, R. Fleury, A. Alu, and A. F. Koenderink, "Drexhage's experiment for sound," Phys. Rev. Lett. **116**, 224301 (2016).
13. M. K. Schmidt, L. G. Helt, C. G. Poulton, and M. J. Steel, "Elastic Purcell effect," Phys. Rev. Lett. **121**, 064301 (2018).
14. G. J. Hornig, K. G. Scheuer, E. B. Dew, R. Zemp, and R. G. DeCorby, "Ultrasound sensing at thermomechanical limits with optomechanical buckled-dome microcavities," Opt. Express **30**(18), 33083-33096 (2022).

15. K. G. Scheuer, F. B. Romero, and R. G. DeCorby, "Spectroscopy of substrate thermal vibrational modes using an optomechanical sensor," in press.
16. B.-B. Li, L. Ou, Y. Lei, and Y.-C. Liu, "Cavity optomechanical sensing," Nanophotonics **10**(11), 2799-2832 (2021).
17. S. Basiri-Esfahani, A. Armin, S. Forstner, and W. P. Bowen, "Precision ultrasound sensing on a chip," Nat. Commun. **10**(1), 132 (2019).
18. A. Prosperetti, "Bubbles," Phys. Fluids **16**(6), 1852-1865 (2004).
19. T. G. Leighton, *The Acoustic Bubble*, Academic Press, London (1997).
20. M. Devaud, T. Hocquet, J.-C. Bacri, and V. Leroy, "The Minnaert bubble: an acoustic approach," Eur. J. Phys. **29**, 1263-1285 (2008).
21. M. P. Brenner, S. Hilgenfeldt, and D. Lohse, "Single-bubble sonoluminescence," Rev. Mod. Phys. **74**, 425-484 (2002).
22. K. S. Suslick, "Sonochemistry," Science **247**, 1439-1445 (1990).
23. I. S. Maksymov and A. D. Greentree, "Coupling light and sound: giant nonlinearities from oscillating bubbles and droplets," Nanophotonics **8**(3), 367-390 (2019).
24. M Minnaert, "XVI. On musical air bubbles and the sounds of running water," Phil. Mag. **6**, 235-248 (1933).
25. W. Lauterborn and T. Kurz, "Physics of bubble oscillations," Rep. Prog. Phys. **73**, 106501 (2010).
26. H. Lamb, *Hydrodynamics* (Dover, New York, 1945), Sec. 275.
27. M. Strasberg, "Gas bubbles as sources of sound in liquids," J. Acoust. Soc. Am. **28**(1), 20-26 (1956).
28. Y. Mao, L. A. Crum, and R. A. Roy, "Nonlinear coupling between the surface and volume modes of an oscillating bubble," J. Acoust. Soc. Am. **98**, 2764-2771 (1995).
29. Z. Zhang, Y. Wang, Y. Amarouchene, R. Boisgard, H. Kellay, A. Wurger, and A. Maali, "Near-field probe of thermal fluctuations of a hemispherical bubble surface," Phys. Rev. Lett. **126**, 174503 (2021).
30. J. L. Flanagan, "Acoustic modes of a hemispherical room," J. Acoust. Soc. Am. **37**(4), 616-618 (1965).
31. J. W. S. Rayleigh, *Theory of Sound*, 2nd ed. (Dover, New York, 1945), Vol. 2, pp. 264-268.
32. D. A. Russell, "Basketballs as spherical acoustic cavities," Am. J. Phys. **78**, 549-554 (2010).
33. M. P. Brenner, S. Hilgenfeldt, D. Lohse, and R. R. Rosales, "Acoustic energy storage in single bubble sonoluminescence," Phys. Rev. Lett. **77**(16), 3467-3470 (1996).
34. T. L. Geers, R. S. Lagumbay, and O. V. Vasilyev, "Acoustic-wave effects in violent bubble collapse," J. Appl. Phys. **112**, 054910 (2012).
35. M. S. Plesset and A. Prosperetti, "Bubble dynamics and cavitation," Ann. Rev. Fluid Mech. **9**, 145-185 (1977).
36. R. Dahan, L. L. Martin, and T. Carmon, "Droplet optomechanics," Optica **3**(2), 175-178 (2016).
37. G. J. Hornig, K. G. Scheuer, and R. G. DeCorby, "Observation of thermal acoustic modes of a droplet coupled to an optomechanical sensor," Appl. Phys. Lett. **123**, 042202 (2023).
38. K. G. Scheuer, F. B. Romero, G. J. Hornig, and R. G. DeCorby, "Ultrasonic spectroscopy of sessile droplets coupled to optomechanical sensors," Lab Chip **23**, 5131-5138 (2023).
39. A. O. Maksimov, "On the volume oscillations of a tethered bubble," J. Sound and Vibration **283**, 915-926 (2005).
40. S. Takahashi, "Properties and characteristics of P(VDF/TrFE) transducers manufactured by a solution casting method for use in the MHz-range ultrasound in air," Ultrasonics **52**(3), 422–426 (2012).
41. U. Ingard and G. L. Lamb, "Effect of a reflecting plane on the power output of sound sources," J. Acoust. Soc. Am. **29**(6), 743-744 (1957).
42. A.-W.El-Sayed and S. Hughes, "Quasinormal-mode theory of elastic Purcell factors and Fano resonances of optomechanical beams," Phys. Rev. Research **2**, 043290 (2020).
43. M. Landi, J. Zhao, W. E. Prather, Y. Wu, and L. Zhang, "Acoustic Purcell effect for enhanced emission," Phys. Rev. Lett. **120**, 114301 (2018).
44. H. Yokoyama, "Physics and device applications of optical microcavities," Science **256**, 66-70 (1992).
45. Y. Q. Yu and Z. Zong, "A study of the internal vibration of a single oscillating bubble," Phys. Fluids **33**, 076106 (2021).
46. A. Hashmi, G. Yu, M. Reilly-Collette, G. Heiman, and J. Xu, "Oscillating bubbles: a versatile tool for lab on a chip applications," Lab Chip **12**, 4216-4227 (2012).
47. I. S. Maksymov, B. Q. H. Nguyen, and S. A. Suslov, "Biochemical sensing using gas bubble oscillations in liquids and adjacent technologies: theory and practical applications," Biosensors **12**, 624 (2022).